\documentstyle[preprint,aps,prl]{revtex}

\begin{document}

\draft

\title{Damping in dilute Bose gases: a mean-field approach} 

\author{ S. Giorgini}

\address{European Centre for Theoretical Studies
in Nuclear Physics and Related Areas
\protect \\
Villa Tambosi, Strada delle Tabarelle 286, I-38050
Villazzano, Italy \protect \\
and Istituto Nazionale di Fisica della Materia, \protect \\
Unit\`a di Trento, I-38050 Povo, Italy}


\maketitle

\begin{abstract}

{\it Damping in a dilute Bose gas is investigated using a 
mean-field approximation which describes the coupled oscillations of condensate
and non-condensate atoms in the collisionless regime. Explicit results 
for both Landau and Beliaev damping rates are given for non-uniform gases.
In the case of uniform systems we obtain results for the damping of phonons 
both at 
zero and finite temperature. The isothermal compressibility of a uniform gas
is also discussed.} 

\end{abstract}


\narrowtext

\section{Introduction}

The low-lying collective excitations of magnetically trapped Bose gases have
been the object of very accurate experimental measurements 
\cite{EXP1,EXP2,JILA}. 
The first experiments \cite{EXP1,EXP2} were carried out at low 
temperature, with approximately all the atoms in the condensate state. 
These experiments showed almost undamped oscillations of the condensate at 
frequencies which have been found in excellent agreement with theoretical 
predictions based on the $T=0$ time-dependent Gross-Pitaevskii equation
\cite{THE}. In more recent experiments \cite{JILA} the study of the low-energy 
collective modes has been extended to higher temperatures where the condensate 
oscillates in the presence of a considerably large fraction of above-condensate
atoms. In this case, evidence is given of large frequency shifts 
with unexpected features and of 
strong damping rates which have not yet been understood theoretically.

On the theoretical side, extensions to finite temperature of the 
Gross-Pitaevskii equation  
have been put forward \cite{GRI,HZG,GPS,MCT,DECB}
and have been very successful
in explaining experimental results on the thermodynamic properties of these 
systems, such as condensate fraction, internal energy, specific heat and 
critical temperature \cite{HZG,GPS,MCT}. These 
mean-field descriptions, however, seem to be inadequate when applied 
to the study of 
collective excitations at finite temperature \cite{HZG,DECB}. 
First of all, the proposed mean-field approximations do not account for 
damping. Secondly, the resonance frequencies are predicted to vary with 
temperature mainly because the number of atoms in the condensate changes, and 
the dependence of the frequency shift upon this quantity is expected to show 
the same behaviour as  
the corresponding $T=0$ dependence   
on the total number $N$ of atoms in the trap. None of the features exhibited
by the experimental data on the frequency shift can be explained using these 
descriptions.

In the presently available finite-$T$ extensions of the Gross-Pitaevskii 
equation it is 
assumed that the condensate oscillates in a bath of thermally excited atoms 
at rest and in thermal equilibrium in the effective mean-field 
potential generated by the average condensate density. 
This assumption is valid if the time scale on which the thermal cloud 
oscillates is much larger than the inverse frequency of the condensate 
oscillations. Since both the 
condensate and the thermal cloud 
vary on comparable time scales of the order of the trap frequency, 
a full dynamic treatment of 
the coupled oscillations of the two clouds should be developed. The  
dynamic effects we are aiming to describe  
are irrelevant in the calculation of the thermodynamic properties of 
the system, for which the standard finite-$T$ mean-field approaches are well
suited. In fact, as pointed out in Refs. \cite{GPS,DGGPS}, the fine details 
of the excitation energies do not affect the thermodynamic behaviour, 
for which what matters is the density of states at a given energy. 

Another problem arises as to whether the appropriate regime to
describe the experimental situation in \cite{JILA} is collisional or 
collisionless. In the collisional regime, which takes place at high 
temperatures and densities, the damping mechanisms are of dissipative type
and the dynamic of the system is described by two-fluid hydrodynamics, 
recently developed for trapped gases in Ref. \cite{ZGN}. 
For very dilute systems at low temperatures the mean free path of the 
elementary excitations becomes comparable with the size of the system
and collisions play a minor role. Damping in this regime is not related
to thermalization processes but to coupling between excitations, and 
can be described in the framework of mean field theories. As suggested 
in Ref. \cite{GWS}, the collisionless regime may be appropriate for 
the JILA experiments \cite{EXP1,JILA}, but probably not for the MIT 
experiments \cite{EXP2}.    

In the collisionless regime and at finite temperature the damping of the 
low-lying collective modes is dominated by Landau damping. The idea that 
this mechanism might be relevant to explain the damping rates in trapped 
Bose gases was first suggested by Liu and Schieve \cite{LS} and has then 
been developed in Refs. \cite{PS,LIU,FSW}. In Ref. \cite{PS} Pitaevskii and 
Stringari have derived, using perturbation theory, an expression for Landau
damping which is applicable to trapped Bose gases, and have shown that, 
applied to the uniform case, it reproduces known results for both the low- and 
high-temperature asymptotic behaviour of the phonon damping.

In this paper we develop a time-dependent mean-field approach based on the 
Popov approximation, which describes the dynamics of a Bose-condensed gas 
in the collisionless regime. We obtain a set of 
coupled equations for the condensate and non-condensate 
components which allow us to calculate damping coefficients. 
In non-uniform gases we derive explicit expressions both for Landau damping, 
which
coincides with the finding of Ref. \cite{PS}, and for Beliaev damping.
In the uniform case we reproduce all known results on the damping 
rates of phonons both at $T=0$ and at finite $T$.  

However, the calculation of the temperature dependence of the frequency shifts 
using the Popov approximation is not reliable. For a uniform system  
the calculation should give, in the long wavelength limit, the correction to
the velocity of zeroth sound due to quantum and thermal fluctuations. 
On the other hand, at $T=0$, the 
velocity of zeroth
sound is directly related to the bulk compressibility of the system,
for which we show that the Popov approximation gives an incorrect result.
By studying the isothermal compressibility we also find that the Popov 
approximation is inconsistent at low temperatures, because it neglects 
fluctuations which are relevant for temperatures smaller than the chemical
potential. Since the same fluctuations are also important in the calculation 
of the velocity of zeroth sound in the low temperature regime, we draw the 
conclusion that a dynamic mean-field description, which gives correct account 
of both   
damping rates and frequency shifts, can only be developed going beyond the 
Popov approximation.     
 
The paper is organized as follows: in section 2 we develop the formalism of 
the time-dependent mean-field approximation. In section 3 we study the 
damping of the oscillations in a non-uniform system obtaining explicit
expressions for the Landau and Beliaev damping. In sections 4-A, 4-B we apply 
the results of section 3 to the uniform case. In section 4-C we investigate
the isothermal compressibility of a uniform gas beyond the Popov approximation.

\section{Time-dependent mean-field approximation}

In the presence of a non-uniform 
external field $V_{ext}({\bf r})$ the grand-canonical Hamiltonian of the 
system has the form  
\begin{eqnarray}
K  \equiv H-\mu N &=& \int\;d{\bf r} \;\psi^{\dagger}({\bf r},t)
\left( -\frac{\hbar^2\nabla^2}
{2m} + V_{ext}({\bf r}) - \mu \right) \psi({\bf r},t) \nonumber \\
&+& \frac{{\rm g}}{2} \int\;d{\bf r} \;\psi^{\dagger}({\bf r},t)
\psi^{\dagger}({\bf r},t)
\psi({\bf r},t)\psi({\bf r},t) 
\label{gch}
\end{eqnarray}
in terms of the 
creation and annihilation particle field operators $\psi^{\dagger}({\bf r},t)$ 
and $\psi({\bf r},t)$.
In the above equation {\rm g} is the interaction coupling constant, which to 
lowest order in the s-wave scattering length $a$ is given by  
${\rm g}=4\pi\hbar^2a/m$. 
The equation of motion for the particle field operator then follows immediately
and reads 
\begin{eqnarray}
i\hbar\frac{\partial}{\partial t}\psi({\bf r},t) &=& \left[\psi({\bf r},t),K
\right] \nonumber \\
&=& \left(-\frac{\hbar^2\nabla^2}{2m}+V_{ext}({\bf r})-\mu\right)
\psi({\bf r},t) + 
{\rm g}\psi^{\dagger}({\bf r},t)\psi({\bf r},t)\psi({\bf r},t) \;\;.
\label{heq}
\end{eqnarray}
According to the usual treatment for Bose systems with broken gauge symmetry we 
define a time-dependent condensate wavefunction $\Phi({\bf r},t)$ \cite{HM}  
\begin{equation}
\Phi({\bf r},t)=\langle\psi({\bf r},t)\rangle \;\;,
\label{bgs}
\end{equation}
which allows us to describe situations where the system is displaced from
equilibrium and the condensate
is oscillating in time.  
The average $\langle ...\rangle$ in equation (\ref{bgs}) is thus intended
to be a non-equilibrium average, while  
time-independent equilibrium averages will be 
indicated in this paper with the symbol $\langle ...\rangle_0$. The particle 
field operator can then be decomposed into a condensate and a non-condensate 
component
\begin{equation}
\psi({\bf r},t)=\Phi({\bf r},t)+\tilde{\psi}({\bf r},t) \;\;,
\label{decomp}
\end{equation}
and the non-condensate term satisfies, by definition, the condition 
$\langle\tilde{\psi}
({\bf r},t)\rangle=0$. By applying the decomposition (\ref{decomp}) to the 
Heisenberg equation (\ref{heq}) the term cubic in the field operators becomes  
(all quantities depend on ${\bf r}$ and $t$)
\begin{eqnarray}
\psi^{\dagger}\psi\psi &=& |\Phi|^2\Phi + 2|\Phi|^2\tilde{\psi} + 
\Phi^2\tilde{\psi}^{\dagger} + 2\Phi\tilde{\psi}^{\dagger}\tilde{\psi} 
\nonumber \\
&+& \Phi^{\ast}\tilde{\psi}\tilde{\psi} + \tilde{\psi}^{\dagger}\tilde{\psi}
\tilde{\psi} \;\;.
\label{cubic}
\end{eqnarray}
We assume that for dilute systems the cubic product of the 
non-condensate operators (last term in eq. (\ref{cubic})) has a negligible
effect on the dynamics of the condensate and we set its average value 
equal to zero:
\begin{equation}
\langle\tilde{\psi}^{\dagger}({\bf r},t)\tilde{\psi}({\bf r},t)\tilde{\psi}
({\bf r},t)\rangle = 0 \;\;.
\label{cubav}
\end{equation}
One thus obtains the following equation for   
the time rate of change of $\langle\psi({\bf r},t)\rangle$  
\begin{eqnarray}
i\hbar\frac{\partial}{\partial t}\Phi({\bf r},t) &=& \left(-\frac{\hbar^2
\nabla^2}{2m}+V_{ext}({\bf r})-\mu\right)\Phi({\bf r},t) + 
{\rm g}|\Phi({\bf r},t)|^2
\Phi({\bf r},t) \nonumber \\
&+& 2{\rm g}\Phi({\bf r},t) \tilde{n}({\bf r},t) 
+ {\rm g}\Phi^{\ast}({\bf r},t) \tilde{m}({\bf r},t)
\;\;,
\label{gpeq}
\end{eqnarray}
where we have introduced the normal and anomalous time-dependent densities
defined, respectively, as 
\begin{eqnarray}
\tilde{n}({\bf r},t) &=& \langle\tilde{\psi}^{\dagger}({\bf r},t)\tilde{\psi}
({\bf r},t)\rangle  \nonumber \\
\tilde{m}({\bf r},t) &=& \langle\tilde{\psi}({\bf r},t)\tilde{\psi}({\bf r},t) 
\rangle  \;\;.
\label{densit}
\end{eqnarray}
The equation of motion (\ref{gpeq}) with $\tilde{n}=\tilde{m}=0$ corresponds 
to the usual $T=0$  
Gross-Pitaevskii
equation for the condensate wavefunction, while its extension including the 
normal and anomalous densities has been already discussed by many authors 
both in the study of thermodynamic properties and of the collective modes  
at finite temperature  
\cite{GRI,HZG,GPS,MCT,DECB,ZGN}. 
The novelty here is that we will treat both 
terms within a time-dependent mean-field approximation holding in the 
collisionless regime which is the object of the present work \cite{PBS}. 

We are interested in the small-amplitude regime in which the condensate is 
only slightly displaced from its stationary value
$\Phi_0({\bf r})=\langle\psi({\bf r})\rangle_0$
\begin{equation}
\Phi({\bf r},t)=\Phi_0({\bf r})+\delta\Phi({\bf r},t) \;\;,
\label{flucphi}
\end{equation}
where $\delta\Phi({\bf r},t)$ is a small fluctuation.
In the same way we consider small fluctuations of the normal and anomalous 
densities
\begin{eqnarray}
\tilde{n}({\bf r},t) &=& \tilde{n}^0({\bf r}) 
+ \delta \tilde{n}({\bf r},t) \nonumber \\
\tilde{m}({\bf r},t) &=& \tilde{m}^0({\bf r}) + \delta \tilde{m}({\bf r},t) 
\label{flucden}
\end{eqnarray}
around their  equilibrium values $\tilde{n}^0({\bf r})=\langle\tilde{\psi}
^{\dagger}({\bf r})\tilde{\psi}({\bf r})\rangle_0$ and $\tilde{m}^0({\bf r})=
\langle\tilde{\psi}({\bf r})\tilde{\psi}({\bf r})\rangle_0$. 
  
In the so-called Popov  approximation the 
effects arising from the equilibrium value of the anomalous density in eq.
(\ref{gpeq}) 
are neglected and the following ansatz is introduced in the mean-field scheme  
\begin{equation}
\tilde{m}^0({\bf r}) = 0 \;\;.
\label{Popov}
\end{equation}    
This approximation has been first introduced by Popov in the study of a uniform 
weakly interacting Bose gas at finite temperature \cite{POP} (for a detailed 
discussion see Ref. \cite{GRI} and \cite{SHI}). More recently the Popov 
approximation has been   
extensively used 
in the study of 
properties 
of magnetically trapped Bose gases at finite temperature 
\cite{GRI,HZG,GPS,DECB,ZGN}. The Popov approximation gives a gapless spectrum
of elementary excitations which at $T=0$ coincides with the well-known 
Bogoliubov 
dispersion relation, while at high $T$ it approaches the finite-temperature 
Hartree-Fock spectrum \cite{HF}.
 
By using the Popov prescription (\ref{Popov}), the real wavefunction 
$\Phi_0({\bf r})$ 
satisfies the
stationary equation 
\begin{equation}
\left(-\frac{\hbar^2\nabla^2}{2m}+V_{ext}({\bf r})-\mu+{\rm g}(n_0({\bf r})+
2\tilde{n}^0({\bf r}))\right)\Phi_0({\bf r}) = 0 \;\;,
\label{statgp}
\end{equation}
where  
$n_0({\bf r})=|\Phi_0({\bf r})|^2$ is the condensate density, while the 
time-dependent equation for $\delta\Phi({\bf r},t)$ is obtained by linearizing 
the equation of motion (\ref{gpeq})
\begin{eqnarray}
i\hbar\frac{\partial}{\partial t}\delta\Phi({\bf r},t) &=& \left( -\frac
{\hbar^2\nabla^2}{2m}+V_{ext}({\bf r})-\mu+2{\rm g} n({\bf r})\right)\delta\Phi
({\bf r},t) 
+ {\rm g}n_0({\bf r})\delta\Phi^{\ast}({\bf r},t) \nonumber \\
&+& 2{\rm g}\Phi_0({\bf r})\delta 
\tilde{n}({\bf r},t) + {\rm g}\Phi_0({\bf r})\delta \tilde{m}({\bf r},t) \;\;,
\label{fluceq}   
\end{eqnarray}
where we have introduced the total equilibrium density 
$n({\bf r})=n_0({\bf r})+\tilde{n}^0({\bf r})$. 

From equation (\ref{fluceq}) one clearly sees that the oscillations of the 
condensate are coupled to the fluctuations 
$\delta\tilde{n}
({\bf r},t)$ and $\delta\tilde{m}({\bf r},t)$ of the normal and anomalous 
densities for which we need the
equations of motion. To this aim it is convenient to express  
the non-condensate operators $\tilde{\psi}$, $\tilde
{\psi}^{\dagger}$ in terms of quasiparticle operators $\alpha_j$, $\alpha_j^
{\dagger}$ by means of the Bogoliubov linear transformations
\begin{eqnarray}
\tilde{\psi}({\bf r},t) &=& \sum_j\left(u_j({\bf r})\alpha_j(t) + v_j^{\ast}
({\bf r})\alpha_j^{\dagger}(t)\right) \nonumber \\
\tilde{\psi}^{\dagger}({\bf r},t) &=& \sum_j\left( u_j^{\ast}({\bf r})
\alpha_j^{\dagger}(t) + v_j({\bf r})\alpha_j(t)\right) \;\;.
\label{bogtrans}
\end{eqnarray}
The normalization condition for the functions $u_j({\bf r})$, $v_j({\bf r})$, 
which ensures that the quasiparticle operators $\alpha_j$, $\alpha_j^{\dagger}$
satisfy Bose commutation relations, reads 
\begin{equation}
\int d{\bf r} \left(u_i^{\ast}({\bf r})u_j({\bf r}) - v_i^{\ast}({\bf r})
v_j({\bf r})\right) = \delta_{ij} \;\;.
\label{norm}
\end{equation}
By using the transformations (\ref{bogtrans}) 
the quantities $\tilde{n}({\bf r},t)$ and $\tilde{m}({\bf r},t)$ can be 
expressed in terms of the normal and anomalous quasiparticle 
distribution functions defined by 
\begin{eqnarray}
f_{ij}(t) &=& \langle\alpha_i^{\dagger}(t)\alpha_j(t)\rangle \nonumber \\
g_{ij}(t) &=& \langle\alpha_i(t)\alpha_j(t)\rangle \;\;.
\label{qpdens}
\end{eqnarray}
The time evolution of these functions is fixed by the following equations of 
motion
\begin{eqnarray}
i\hbar\frac{\partial}{\partial t}f_{ij}(t) &=& \left< \biggl[ 
\alpha_i^{\dagger}
(t)\alpha_j(t),K \biggr] \right> \nonumber \\
i\hbar\frac{\partial}{\partial t}g_{ij}(t) &=& \left< \biggl[ \alpha_i
(t)\alpha_j(t),K \biggr] \right>  \;\;.
\label{fgeq}
\end{eqnarray}
In order to calculate the commutators of eq. 
(\ref{fgeq}) one notices that, after substituting the decomposition 
(\ref{decomp})
of the particle field operator into the Hamiltonian (\ref{gch}), 
only the terms quadratic and 
quartic in the non-condensate operators     
$\tilde
{\psi}$, $\tilde{\psi}^{\dagger}$ give non-vanishing contributions,  
because, 
according 
to eq. (\ref{cubav}), we set to zero all averages of cubic products of the  
non-condensate operators. The terms in the grand-canonical 
Hamiltonian relevant for the calculation of the commutators are thus given by
\begin{eqnarray}
K_2+K_4 &=& \int d^3{\bf r} \Biggl[ \tilde{\psi}^{\dagger}({\bf r},t)\Biggl(
-\frac{\hbar^2\nabla^2}{2m}+V_{ext}({\bf r})-\mu \Biggr)\tilde{\psi}({\bf r},t)
+ 2{\rm g}|\Phi({\bf r},t)|^2\tilde{\psi}^{\dagger}({\bf r},t)
\tilde{\psi}({\bf r},t)
\Biggr. \nonumber \\
&+& \Biggl.
\frac{{\rm g}}{2}\Phi^2({\bf r},t)\tilde{\psi}^{\dagger}({\bf r},t)
\tilde{\psi}^{\dagger}({\bf r},t)    
+\frac{{\rm g}}{2}\Phi^{\ast\;2}({\bf r},t)\tilde{\psi}({\bf r},t)\tilde
{\psi}({\bf r},t) \Biggr] \\        
\label{k2k4}
&+&\frac{{\rm g}}{2}\int d^3{\bf r} \;\tilde{\psi}^{\dagger}
({\bf r},t)\tilde{\psi}^{\dagger}({\bf r},t)\tilde{\psi}({\bf r},t)
\tilde{\psi}({\bf r},t) \nonumber \;\;,
\end{eqnarray}
where we have indicated with $K_2$ the term quadratic in 
$\tilde{\psi}$, $\tilde{\psi}^{\dagger}$, 
while $K_4$ is the term quartic in these operators. By expanding $K_2$ up 
to terms linear in the fluctuations $\delta\Phi({\bf r},t)$ we can rewrite 
it as the sum $K_2=K_2^{(0)}+K_2^{(1)}$, where $K_2^{(0)}$ does not 
contain the fluctuations of the condensate
\begin{eqnarray} 
K_2^{(0)} &=& \int d^3{\bf r} \Biggl[ \tilde{\psi}({\bf r},t)
\Biggl( -\frac{\hbar^2\nabla^2}
{2m} + V_{ext}({\bf r}) - \mu + 
2{\rm g}n_0({\bf r})\Biggr)\tilde{\psi}({\bf r},t)
\Biggr. \nonumber \\
&+& \Biggl. \frac{{\rm g}}{2}n_0({\bf r}) 
\left(\tilde{\psi}^{\dagger}({\bf r},t)
\tilde{\psi}
^{\dagger}({\bf r},t) + \tilde{\psi}({\bf r},t)\tilde{\psi}({\bf r},t)
\right) \Biggr] \;\;,
\label{k20}
\end{eqnarray}
while $K_2^{(1)}$ is linear in the fluctuations $\delta\Phi$
\begin{eqnarray}
K_2^{(1)} &=& \int d^3{\bf r} \biggl[ 2{\rm g}\Phi_0({\bf r})
\biggl(\delta\Phi({\bf r},
t)+\delta\Phi^{\ast}({\bf r},t)\biggr) \tilde{\psi}^{\dagger}({\bf r},t)
\tilde{\psi}({\bf r},t) \biggr. \nonumber \\
&+& \biggl. {\rm g}\Phi_0({\bf r}) 
\biggl(\delta\Phi({\bf r},t)\tilde{\psi}^{\dagger}
({\bf r},t)\tilde{\psi}^{\dagger}({\bf r},t) + \delta\Phi^{\ast}({\bf r},t) 
\tilde{\psi}({\bf r},t)\tilde{\psi}({\bf r},t)\biggr) \biggr] \;\;.
\label{k21}
\end{eqnarray}
In the quartic term $K_4$ we first use 
the mean field decomposition (all quantities depend on ${\bf r}$ and $t$)
\begin{equation}
\tilde{\psi}^{\dagger}\tilde{\psi}^{\dagger}\tilde{\psi}\tilde{\psi} = 
4\tilde{n}\tilde{\psi}^{\dagger}\tilde{\psi}+\tilde{m}\tilde{\psi}^{\dagger}
\tilde{\psi}^{\dagger} +\tilde{m}^{\ast}\tilde{\psi}\tilde{\psi} \;\;,
\label{decomp1}
\end{equation}
and then we expand the resulting expression up to linear terms in the 
fluctuations $\delta\tilde{n}$ and $\delta\tilde{m}$ thereby obtaining 
\begin{eqnarray}
K_4 = K_4^{(0)}+K_4^{(1)} &=& 2{\rm g}\int d^3{\bf r} \;\tilde{n}^0({\bf r})
\tilde{\psi}^{\dagger}({\bf r},t)\tilde{\psi}({\bf r},t) 
+ \frac{{\rm g}}{2}\int d^3{\bf r} \biggl[ 4\delta\tilde{n}({\bf r},t)
\tilde{\psi}^{\dagger}
({\bf r},t)\tilde{\psi}({\bf r},t) \biggr. \nonumber \\ 
&+& \biggl. \biggl(\delta\tilde{m}
({\bf r},t)\tilde{\psi}^{\dagger}({\bf r},t)\tilde{\psi}^{\dagger}({\bf r},t)
+\delta\tilde{m}^{\ast}({\bf r},t)\tilde{\psi}({\bf r},t)\tilde{\psi}
({\bf r},t)\biggr) \biggr] \;\;,
\label{k401}
\end{eqnarray}
where we have made use of the Popov ansatz (\ref{Popov}). 
In the above equation 
$K_4^{(0)}$ is the zeroth order term, while $K_4^{(1)}$ is linear in the 
fluctuations $\delta\tilde{n}$ and 
$\delta\tilde{m}$. 
In the equations of motion (\ref{fgeq})   
the commutators of $\alpha_i^{\dagger}\alpha_j$ and
$\alpha_i\alpha_j$ with $K_2^{(1)}$ yield, in the small-amplitude regime, 
the coupling to the fluctuations of the condensate, whereas the 
commutators with $K_4^{(1)}$ give the coupling to the fluctuations 
of the normal and anomalous particle densities. 
If the density of the non-condensate particles is much smaller than the 
density $n_0$ of the condensate, the coupling to the condensate is more 
important than the coupling to $\delta\tilde{n}$ and $\delta\tilde{m}$
and we can consequently neglect the contributions to the commutators arising
from the term $K_4^{(1)}$.  

One can easily show that the operator $K_2^{(0)}+K_4^{(0)}$ is diagonal in the 
quasiparticle operators $\alpha_j$, $\alpha_j^{\dagger}$ if 
the functions $u_j$ and $v_j$ satisfy
the coupled Bogoliubov equations
\begin{eqnarray}
{\cal L}u_j({\bf r})+{\rm g}n_0({\bf r})v_j({\bf r}) &=& \epsilon_ju_j({\bf r})
\nonumber \\
{\cal L}v_j({\bf r})+{\rm g}n_0({\bf r})u_j({\bf r}) &=& - \epsilon_j
v_j({\bf r}) \;\;, 
\label{bogeqs}
\end{eqnarray}
where we have introduced the hermitian operator
\begin{equation}
{\cal L} = - \frac{\hbar^2\nabla^2}{2m} + V_{ext}({\bf r}) - \mu
+ 2{\rm g}n({\bf r}) \;\;.
\label{Lop}
\end{equation}
As a consequence the relevant terms in the Hamiltonian become  
\begin{equation}
K_2+K_4=\sum_j\epsilon_j\alpha_j^{\dagger}\alpha_j + K_2^{(1)} \;\;,
\label{relham}
\end{equation}
where the quasiparticle energies $\epsilon_j$ are obtained from the solutions
of the Bogoliubov equations (\ref{bogeqs}).

The commutators in the equations of motion (\ref{fgeq}) can be now calculated
straightforwardly. To lowest order in the fluctuations one gets  
\begin{eqnarray}
i\hbar\frac{\partial}{\partial t}f_{ij}(t) &=& (\epsilon_j-\epsilon_i)f_{ij}(t)
+ 2{\rm g}(f_i^0-f_j^0)\int d^3{\bf r} \;\Phi_0({\bf r}) 
\biggl[ \biggl(\delta\Phi({\bf r},t)+\delta\Phi^{\ast}({\bf r},t)\biggr) 
\times \biggr.
\nonumber \\
&& \biggl.
\biggl(u_i({\bf r})u_j^{\ast}({\bf r})+v_i({\bf r})v_j^{\ast}({\bf r})\biggr)
+\delta\Phi({\bf r},t)v_i({\bf r})u_j^{\ast}({\bf r}) + \delta\Phi^{\ast}
({\bf r},t)u_i({\bf r})v_j^{\ast}({\bf r}) \biggr] 
\label{feq}
\end{eqnarray}
for the time evolution of $f_{ij}$,
while the equation of motion for $g_{ij}$ is given by
\begin{eqnarray}
i\hbar\frac{\partial}{\partial t}g_{ij}(t) &=& (\epsilon_j+\epsilon_i)g_{ij}(t)
+ 2{\rm g}(1+f_i^0+f_j^0)\int d^3{\bf r} \;\Phi_0({\bf r}) 
\biggl[ \biggl(\delta\Phi({\bf r},t)+\delta\Phi^{\ast}({\bf r},t)\biggr) 
\times \biggr.
\nonumber \\
&& \biggl.
\biggl(u_i^{\ast}({\bf r})v_j^{\ast}({\bf r})
+v_i^{\ast}({\bf r})u_j^{\ast}({\bf r})\biggr)
+\delta\Phi({\bf r},t)u_i^{\ast}({\bf r})u_j^{\ast}({\bf r}) + \delta\Phi^{\ast}
({\bf r},t)v_i^{\ast}({\bf r})v_j^{\ast}({\bf r}) \biggr] \;\;.
\label{geq}
\end{eqnarray}
In the equations (\ref{feq})-(\ref{geq}) $f_j^0$ is the  
equilibrium density of quasiparticles
\begin{equation}
f_j^0 = \langle\alpha_j^{\dagger}\alpha_j\rangle = \left(\exp(\beta\epsilon_j)
-1\right)^{-1}
\label{qpdist}
\end{equation}
in terms of which the non-condensate particle density, at equilibrium, 
is written as
\begin{equation}
\tilde{n}^0({\bf r}) = \sum_j \left[ (|u_j({\bf r})|^2+|v_j({\bf r})|^2) f_j^0
+ |v_j({\bf r})|^2 \right] \;\;.
\label{pdens}
\end{equation}
The last term in (\ref{pdens}) accounts at $T=0$ for the quantum 
depletion of the condensate.
The fluctuations $\delta\tilde{n}$ and $\delta\tilde{m}$ of 
the normal and anomalous particle densities can be straightforwardly expressed 
in terms of 
$f_{ij}(t)$ and $g_{ij}(t)$ and the equation (\ref{fluceq}) for the 
oscillations of the condensate takes the final form 
\begin{eqnarray}
i\hbar\frac{\partial}{\partial t}\delta\Phi({\bf r},t) &=& \left( -\frac
{\hbar^2\nabla^2}{2m}+V_{ext}({\bf r})-\mu+2{\rm g} n({\bf r})\right)\delta\Phi
({\bf r},t) 
+ {\rm g}n_0({\bf r})\delta\Phi^{\ast}({\bf r},t)  \nonumber \\
&+& {\rm g}\Phi_0({\bf r}) \sum_{ij} \Biggl[ 
2\biggl(u_i^{\ast}({\bf r})u_j({\bf r})+v_i^{\ast}
({\bf r})v_j({\bf r})+v_i^{\ast}({\bf r})u_j({\bf r})\biggr) f_{ij}(t)
\Biggr.  
\label{fluceq1}
\\
&+& \Biggl. 
\biggl(2v_i({\bf r})u_j({\bf r})+u_i({\bf r})
u_j({\bf r})\biggr)g_{ij}(t) + \biggl(2u_i^{\ast}({\bf r})v_j^{\ast}({\bf r})
+v_i^{\ast}({\bf r})v_j^{\ast}({\bf r})\biggr) g_{ij}^{\ast}(t) \Biggr] \;\;.
\nonumber
\end{eqnarray}
The equations of motion    
(\ref{feq})-(\ref{geq}) and (\ref{fluceq1}) describe the 
small-amplitude coupled oscillations of the condensate and 
non-condensate particles in the collisionless regime. They represent the main 
result of this section. 

\section{Damping of the oscillations}

Let us suppose that the condensate oscillates with frequency $\omega$
\begin{equation}
\delta\Phi({\bf r},t) = \delta\Phi_1({\bf r}) e^{-i\omega t} 
\;\;\;\;\;\;\;
\delta\Phi^{\ast}({\bf r},t) = \delta\Phi_2({\bf r}) e^{-i\omega t} \;\;.
\label{fourier}
\end{equation}
Through the coupling term in the equations of motion (\ref{feq})-(\ref{geq}) 
the fluctuations of the condensate act as a time-dependent external 
drive inducing oscillations in $f_{ij}$ and $g_{ij}$. The Fourier component
of $f_{ij}$ at the driving frequency $\omega$ is given by
\begin{eqnarray}
f_{ij}(\omega) = 2{\rm g} \frac{f_i^0-f_j^0}{\hbar\omega+(\epsilon_i-\epsilon_j)
+i0} && \int d^3{\bf r} \;\Phi_0\biggl[\delta\Phi_1\left(u_iu_j^{\ast}+v_iv_j
^{\ast}+v_iu_j^{\ast}\right) \biggr. \nonumber \\
&+& \biggl. \delta\Phi_2\left(u_iu_j^{\ast}+v_iv_j^{\ast}+u_iv_j^{\ast}\right)
\biggr] \;\;,
\label{fomega}
\end{eqnarray}
where the frequency $\omega$ has been chosen with an infinitesimally small 
component on the positive imaginary axis. Analogously, for the component 
of $g_{ij}$ 
oscillating at the frequency $\omega$ one finds 
\begin{eqnarray}
g_{ij}(\omega) = 2{\rm g} \frac{1+f_i^0+f_j^0}
{\hbar\omega-(\epsilon_i+\epsilon_j)
+i0} && \int d^3{\bf r} \;\Phi_0\biggl[\delta\Phi_1\left(u_i^{\ast}v_j^{\ast}
+v_i^{\ast}u_j
^{\ast}+u_i^{\ast}u_j^{\ast}\right) \biggr. \nonumber \\
&+& \biggl. \delta\Phi_2\left(u_i^{\ast}v_j^{\ast}+v_i^{\ast}u_j^{\ast}
+v_i^{\ast}v_j^{\ast}\right)
\biggr] \;\;.
\label{gomega}
\end{eqnarray}
If one neglects in equation (\ref{fluceq1}) the 
coupling terms involving the fluctuations of the non-condensate particles, 
the condensate components $\delta\Phi_1^0$, $\delta\Phi_2^0$ satisfy the 
unperturbed RPA equation \cite{HZG}
\begin{eqnarray}
\left( \begin{array}{cc} {\cal L}        & {\rm g}n_0({\bf r}) \\
                   -{\rm g}n_0({\bf r})  & -{\cal L}
\end{array} \right)
\left( \begin{array}{c} \delta\Phi_1^0({\bf r})  \\
                        \delta\Phi_2^0({\bf r})
\end{array} \right)
= \hbar\omega_0 \left( \begin{array}{c} \delta\Phi_1^0({\bf r})  \\
                                        \delta\Phi_2^0({\bf r})
\end{array} \right)  \;\;,
\label{rpa0}
\end{eqnarray}
and the normalization condition
\begin{equation}
\int d^3{\bf r} \left( |\delta\Phi_1^0|^2 - |\delta\Phi_2^0|^2 \right) = 1 \;\;,
\label{rpanorm0}
\end{equation} 
where $\omega_0$ is the unperturbed RPA eigenfrequency of the mode.
We treat the coupling terms in equation (\ref{fluceq1}) as small perturbations 
and we write the solution of
the normal mode as
\begin{eqnarray}
\left( \begin{array}{c} \delta\Phi_1 \\
                        \delta\Phi_2
\end{array} \right)
= \left( \begin{array}{c} \delta\Phi_1^0 \\
                          \delta\Phi_2^0
\end{array} \right)
+ \left( \begin{array}{c} \delta\Phi_1^{\prime} \\
                          \delta\Phi_2^{\prime}
\end{array} \right) \;\;,
\label{pervec}
\end{eqnarray}
and the perturbed eigenfrequency as
\begin{equation}
\hbar\omega=\hbar\omega_0 + \delta E - i\gamma \;\;.
\label{perfre}
\end{equation}
In eq. (\ref{perfre}) 
$\delta E$ represents the shift in the real part of the frequency and
$\gamma$ is the damping coefficient, while the eigenvector correction 
in eq. (\ref{pervec}) is chosen  
to be orthogonal to the unperturbed eigenvector
\begin{equation}
\int d^3{\bf r} \left(\delta\Phi_1^{\prime\ast}\delta\Phi_1^0 - 
\delta\Phi_2^{\prime\ast}\delta\Phi_2^0\right) = 0 \;\;.
\label{ortcon}
\end{equation}
Starting from the equations for the perturbed components $\delta\Phi_1$
and $\delta\Phi_2$ [eq. (\ref{fluceq1}) and its 
complex conjugate] we multiply the first equation 
by $\delta\Phi_1^{\ast}$ 
and the 
latter by $\delta\Phi_2^{\ast}$, take the difference of the two equations 
and finally integrate over space. By using (\ref{rpanorm0}) and (\ref{ortcon}) 
we get the 
following relation for the perturbed eigenfrequency   
\begin{eqnarray}
\hbar\omega 
= \hbar\omega_0 
&+&4{\rm g}^2\sum_{ij} (f_i^0-f_j^0)\frac{|A_{ij}|^2}{\hbar\omega_0+(\epsilon_i-
\epsilon_j)+i0} \nonumber \\
&+&2{\rm g}^2\sum_{ij} (1+f_i^0+f_j^0)\left( \frac{|B_{ij}|^2}{\hbar\omega_0-
(\epsilon_i+\epsilon_j)+i0} - \frac{|\tilde{B}_{ij}|^2}{\hbar\omega_0+
(\epsilon_i+\epsilon_j)+i0}\right) ,
\label{perrpa}
\end{eqnarray}
where the matrix elements $A_{ij}$, $B_{ij}$ and $\tilde{B}_{ij}$ are
respectively given by
\begin{eqnarray}
A_{ij} &=& \int d^3{\bf r} \;\Phi_0 \left[\delta\Phi_1^0 (u_iu_j^{\ast}+
v_iv_j^{\ast}+v_iu_j^{\ast})+\delta\Phi_2^0 (u_iu_j^{\ast}+v_iv_j^{\ast}
+u_iv_j^{\ast})\right]
\nonumber \\
B_{ij} &=& \int d^3{\bf r} \;\Phi_0 \left[\delta\Phi_1^0 (u_i^{\ast}v_j^{\ast}+
v_i^{\ast}u_j^{\ast}+u_i^{\ast}u_j^{\ast})
+\delta\Phi_2^0 (u_i^{\ast}v_j^{\ast}+v_i^{\ast}u_j^{\ast}
+v_i^{\ast}v_j^{\ast})\right]
\label{matrix}
\\
\tilde{B}_{ij} &=& \int d^3{\bf r} \;\Phi_0 \left[\delta\Phi_1^0 (u_iv_j+
v_iu_j+v_iv_j)+\delta\Phi_2^0 (u_iv_j+v_iu_j
+u_iu_j)\right] \;\;. \nonumber
\end{eqnarray}

The imaginary part of the right hand side of eq. (\ref{perrpa}) gives  
the damping coefficient $\gamma$. There are two distinct contributions
to $\gamma$: one arises from the process of one
quantum of oscillation $\hbar\omega_0$ being absorbed by a thermal excitation
with energy $\epsilon_i$ which is turned into another thermal excitation 
with energy $\epsilon_j=\epsilon_i+\hbar\omega_0$. This mechanism is 
known as Landau ($L$) damping and is given by the imaginary part of the second 
term on the right hand side of eq. (\ref{perrpa}) 
\begin{equation}
\gamma_L=4\pi g^2\sum_{ij} |A_{ij}|^2 \left(f_i^0-f_j^0\right)
\delta(\hbar\omega_0+\epsilon_i-\epsilon_j) \;\;.
\label{Landam}
\end{equation}
The above result coincides with the finding of Ref. \cite{PS} obtained by 
direct use of 
perturbation theory.
A different mechanism of damping comes from the process  
where a quantum of 
oscillation $\hbar\omega_0$ is absorbed and two excitations with energies 
$\epsilon_i+\epsilon_j=\hbar\omega_0$ are created. The damping associated with 
the decay of an elementary excitation into a pair of excitations, which
has been studied by Beliaev in the case of uniform Bose superfluids 
\cite{BEL}, is 
present also at $T=0$, but is not active for the lowest energy modes in the 
case of a trapping potential because of the discretization of levels.
The Beliaev ($B$) damping is given by the imaginary part of the first term 
in brackets on the right hand side of 
eq. (\ref{perrpa}) and reads
\begin{equation}
\gamma_B=2\pi g^2\sum_{ij} |B_{ij}|^2 \left(1+f_i^0+f_j^0\right)
\delta(\hbar\omega_0-\epsilon_i-\epsilon_j) \;\;.
\label{Beldam}
\end{equation}
The total damping coefficient is the sum of the two contributions 
$\gamma=\gamma_L+\gamma_B$, the Beliaev damping 
becomes dominant at low temperatures ($k_BT\ll\hbar\omega_0$), while 
the Landau damping becomes dominant in the opposite regime of temperatures 
($k_BT\gg\hbar\omega_0$). 

In the case of magnetically trapped Bose gases result (\ref{Landam}) 
and (\ref{Beldam})
give  the damping coefficient $\gamma$ both at low and high temperatures 
in the collisionless regime. This quantity 
has been measured at JILA over a wide range of temperatures for the $m=0$ 
and $m=2$ modes \cite{JILA}, revealing a very strong temperature dependence.  
To carry out the 
calculation at a given temperature $T$ one must start from the unperturbed 
condensate eigenfrequency 
with the 
required symmetry, obtained from eq. (\ref{rpa0}), and the elementary 
excitation energies $\epsilon_j$ calculated from eqs. (\ref{bogeqs}).
One has then to calculate the matrix elements $A_{ij}$ and $B_{ij}$ (see eq. 
(\ref{matrix})) and then carry out the double summation in (\ref{Landam}) 
and (\ref{Beldam}). The explicit calculation of $\gamma$ in 
magnetically trapped gases will be the object of a future work.

By analyzing the real part of eq. (\ref{perrpa}) one can calculate the 
eigenfrequency shift $\delta E$ in the Popov approximation. However,   
for temperatures $k_BT\leq\mu$ 
the effects arising from the 
equilibrium anomalous density
$\tilde{m}^0({\bf r})$ are important for the calculation of the frequency 
shift and can not be neglected. In section 4-C we will present a calculation 
of the velocity of sound at $T=0$ for a uniform gas and 
it will be shown that the 
inclusion of $\tilde{m}^0$ is crucial to obtain the correct result.  
From this finding we conclude that a 
reliable calculation of the frequency shift $\delta E$ should be based on 
a more accurate dynamic theory, beyond the simple Popov ansatz (\ref{Popov}).

\section{Uniform Bose-condensed gases}

In this section we apply the results (\ref{Landam})-(\ref{Beldam}) to uniform
Bose gases for which we reproduce well known results for the damping of 
phonons, both at finite and zero temperature,  
in the collisionless regime.

For homogeneous systems the stationary condensate wavefunction is constant
throughout space 
$\Phi_0=\sqrt{n_0}$, while the condensate fluctuations and the 
excitations are described by plane wave functions
\begin{equation}
\delta\Phi_1({\bf r})=\frac{u_q}{\sqrt{V}}e^{i{\bf q}\cdot{\bf r}} \;\;\;
\delta\Phi_2({\bf r})=\frac{v_q}{\sqrt{V}}e^{i{\bf q}\cdot{\bf r}} \;\;\;
u_{\bf p}({\bf r})=\frac{u_p}{\sqrt{V}}e^{i{\bf p}\cdot{\bf r}} \;\;\;
v_{\bf p}({\bf r})=\frac{v_p}{\sqrt{V}}e^{i{\bf p}\cdot{\bf r}} \;\;,
\label{plwave}
\end{equation}
where $u_p$ and $v_p$ are real functions defined through the usual Bogoliubov
relations 
\begin{eqnarray}
&& u_p^2=1+v_p^2=\frac{(\epsilon_p^2+{\rm g}^2n_0^2)^{1/2}+\epsilon_p}
{2\epsilon_p}
\nonumber \\
&& u_pv_p = - \frac{{\rm g}n_0}{2\epsilon_p} \;\;,
\label{uv}
\end{eqnarray}
and $\epsilon_p$ is the energy of the elementary excitations as obtained 
from the Bogoliubov
equations (\ref{bogeqs})
\begin{equation}
\epsilon_p = \left(\left(\frac{p^2}{2m}+{\rm g}n_0\right)^2
-{\rm g}^2n_0^2\right)^{1/2}
\;\;.
\label{excen}
\end{equation}
In these equations $n_0\equiv n_0(T)$ is the condensate density at the given 
temperature 
$T$ obtained from the self-consistent calculation in thermodynamic 
equilibrium (see Ref. \cite{GPS}).   

\subsection{Thermal regime $\hbar\omega_0\ll k_BT$}

We consider the long-wavelength limit for the fluctuations of the condensate
\begin{equation}
\hbar\omega_0\equiv\epsilon_q\simeq cq \;\;,
\label{con}
\end{equation}
where $c=\sqrt{{\rm g}n_0/m}$ 
is the velocity of sound calculated at temperature $T$. 
In this limit the $u$ and $v$
functions associated with the oscillations of the condensate can be expanded 
as 
\begin{eqnarray}
u_q &\simeq& \left(\frac{mc^2}{2\epsilon_q}\right)^{1/2}
+\frac{1}{2}\left(\frac{\epsilon_q}{2mc^2}
\right)^{1/2} \nonumber \\
v_q &\simeq& - \left(\frac{mc^2}{2\epsilon_q}\right)^{1/2}
+\frac{1}{2}\left(\frac{\epsilon_q}{2mc^2}
\right)^{1/2}  \;\;.
\label{exp}
\end{eqnarray}
As it has been shown in Ref. \cite{PS}, using the above expansion for $u_q$ 
and $v_q$ 
one obtains the following low-$q$ behaviour for the matrix elements 
$A_{{\bf p}{\bf p}'}$ defined in 
(\ref{matrix}) 
\begin{equation}
A_{{\bf p}{\bf p}'}=\delta_{{\bf p}',{\bf p}+{\bf q}} 
\frac{\sqrt{n_0}}{\sqrt{V}}
\left(\frac{\epsilon_q}{2mc^2}\right)^{1/2} \left( u_p^2+v_p^2+u_pv_p
-\frac{v_g}{c}cos\theta\frac{2u_p^2v_p^2}{u_p^2+v_p^2}\right) \;\;,
\label{matrix1}
\end{equation}
where $\delta_{{\bf p}{\bf p}'}$ is the Kronecker symbol, $\theta$ is 
the angle formed between the directions of ${\bf p}$ and ${\bf q}$, and
$v_g=\partial\epsilon_p/\partial p$ is the group velocity of the 
excitations.
By introducing the above expansion 
into eq. (\ref{Landam}) and carrying out the integration over $\theta$ one 
finds the result (see Ref. \cite{PS})
\begin{equation}
\frac{\gamma}{\epsilon_q}\simeq\frac{\gamma_L}{\epsilon_q}
= \left(a^3n_0\right)^{1/2} F(\tau) \;\;,
\label{Landam1}
\end{equation}
where $\tau=k_BT/mc^2$ is a reduced temperature and 
$F(\tau)$ is the dimensionless function
\begin{equation}
F(\tau)=8\sqrt{\pi}\int dx \left(e^x-e^{-x}\right)^{-2} 
\left(1-\frac{1}{2u}-\frac{1}{2u^2}\right)^2 \;\;,
\label{Ftau}
\end{equation}
where we have introduced the quantity $u=\sqrt{1+4\tau^2x^2}$.

For temperatures $\tau\gg1$ the function $F$ takes the asymptotic limit 
$F\rightarrow 3\pi^{3/2}\tau/4$ and the damping coefficient is given by 
\begin{equation}
\frac{\gamma}{\epsilon_q}=\frac{3\pi}{8}\frac{k_BTa}{\hbar c} \;\;.
\label{kond}
\end{equation}
This regime of temperatures has been first investigated by Sz\'{e}pfalusy and 
Kondor \cite{SK}. 
As discussed in Ref. \cite{PS}, the numerical coefficient in (\ref{kond}) 
coincides with the one obtained in \cite{SHI}, while it 
slightly differs from the one of \cite{SK}.

In the opposite regime of temperatures, $\tau\ll 1$, one finds the asymptotic 
limit $F\rightarrow 3\pi^{9/2}\tau^4/5$ and one recovers the well known 
result for the phonon damping \cite{HM,AK,LP,PS} 
\begin{equation}
\frac{\gamma}{\epsilon_q}=\frac{3\pi^3}{8}\frac{(k_BT)^4}{mn\hbar^3c^5} \;\;.
\label{khal}
\end{equation}

In Fig. 1 the dimensionless function $F(\tau)$ is plotted as a function of
$\tau$ together with its asymptotic behaviour both at small and large  
$\tau$'s. One can see that $F$ departs rather soon from the low-temperature 
$\tau^4$ behaviour, while it approaches the high-temperature  
linear law very slowly.

\subsection{Quantum regime $\hbar\omega_0\gg k_BT$}

At $T=0$ the damping of the long-wavelength fluctuations of the condensate 
with energy (\ref{con}) is obtained through result
(\ref{Beldam}) with $f_j^0=0$. For uniform gases the matrix elements 
$B_{{\bf p}{\bf p}'}$ entering eq. (\ref{Beldam}) read
\begin{equation}
B_{{\bf p}{\bf p}'} = \delta_{{\bf p}',-{\bf p}
+{\bf q}}\frac{\sqrt{n_0}}
{\sqrt{V}}\biggl(u_q\bigl(u_pv_{p-q}+v_pu_{p-q}+u_pu_{p-q}\bigr) + v_q
\bigl(u_pv_{p-q}+v_pu_{p-q}+v_pv_{p-q}\bigr)\biggr) \;\;.
\label{matrix2}
\end{equation}
In the Beliaev damping mechanism the momenta of the three  
excitations involved in the process 
are comparable, $q\simeq p\simeq |{\bf p}-{\bf q}|$.
For $p\ll mc$ one can use the following expansions for the excitation 
energies 
$\epsilon_p$ 
and the functions $u_p$
and $v_p$
\begin{eqnarray}
\epsilon_p &\simeq& cp + \frac{p^3}{8m^2c} \nonumber \\
u_p &\simeq& \left(\frac{mc}{2p}\right)^{1/2} + \frac{1}{2}\left(\frac{p}
{2mc}\right)^{1/2} + \frac{1}{8}\left(\frac{p}{2mc}\right)^{3/2} 
-\frac{1}{8}\left(\frac{p}{2mc}\right)^{5/2}  
\label{exp1} \\
v_p &\simeq& - \left(\frac{mc}{2p}\right)^{1/2} + \frac{1}{2}\left(\frac{p}
{2mc}\right)^{1/2} - \frac{1}{8}\left(\frac{p}{2mc}\right)^{3/2} 
-\frac{1}{8}\left(\frac{p}{2mc}\right)^{5/2} \;\;. \nonumber 
\end{eqnarray}
If one substitutes the above expressions for all the $u$'s and $v$'s in
(\ref{matrix2}) and makes use of the condition for energy conservation  
\begin{equation}
|{\bf p}-{\bf q}|+p=q+\frac{1}{8m^2c^2}\left(q^3+p^3+|{\bf p}-{\bf q}|^3 
\right) \;\;,
\label{encons}
\end{equation}
after long but straightforward algebra one gets the result 
\begin{equation}
B_{{\bf p}{\bf p}'}=\delta_{{\bf p}',-{\bf p}+{\bf q}}\frac{\sqrt{n_0}}
{\sqrt{V}} \frac{3}{4\sqrt{2}}\frac{(qp|{\bf p}-{\bf q}|)^{1/2}}{(mc)^{3/2}} 
\;\;.
\label{matrix3}
\end{equation}
Once the low-energy behaviour of the matrix elements of $B$ 
has been obtained, the 
calculation of the damping coefficient follows directly
\begin{equation}
\gamma\simeq\gamma_B =\frac{3q^5}{640\pi\hbar^3mn} \;\;,
\label{Beldam1}
\end{equation} 
and coincides with the well known result first obtained by Beliaev using 
diagrammatic techniques \cite{BEL,N1} (for a review of the second-order 
Beliaev approximation at $T=0$ and its extension to finite temperature 
see Ref. \cite{SHI}).

\subsection{Bulk compressibility}

In this section we calculate the isothermal inverse compressibility 
$\kappa_T^{-1}=n^2(\partial\mu/\partial n)_{V,T}$. In a uniform system 
at $T=0$ it is fixed  
by the velocity of sound through the relation 
\begin{equation}
\kappa_{T=0}^{-1}=\rho\left(\frac{\partial P}{\partial\rho}\right)_{N,T=0}=
\rho c^2(T=0) \;\;,
\label{invcomp}
\end{equation}
where $\rho=mn$ is the mass density.   
We will show that the Popov approximation gives an incorrect result for the 
inverse 
compressibility in the low-temperature regime and that one must go beyond 
this approximation in order to obtain the low-$T$ behaviour of 
$\kappa_T^{-1}$.

Without making the Popov assumption (\ref{Popov}) and including the term 
proportional to $\tilde{m}^0({\bf r})$, the stationary equation 
for the real wavefunction $\Phi_0({\bf r})$ becomes
\begin{equation}
\left(-\frac{\hbar^2\nabla^2}{2m}+V_{ext}({\bf r})-\mu+{\rm g}(n_0({\bf r})+
2\tilde{n}^0({\bf r})+\tilde{m}^0({\bf r}))\right)\Phi_0({\bf r}) = 0 \;\;,
\label{statgp1} 
\end{equation}
instead of eq. 
(\ref{statgp}).
In terms of the quasiparticles operators defined in 
(\ref{bogtrans}) the anomalous density at equilibrium is given by
\begin{equation}
\tilde{m}^0({\bf r}) = \sum_j\biggl[ 2u_j({\bf r})v_j^{\ast}({\bf r})f_j^0 +
u_j({\bf r})v_j^{\ast}({\bf r}) \biggr] \;\;.
\label{andens}
\end{equation}
For a uniform system the above equation (\ref{statgp1}) fixes the value of 
the chemical potential 
\begin{equation}
\mu={\rm g}n + {\rm g}(\tilde{n}^0+\tilde{m}^0) \;\;
\label{chempot}
\end{equation}
which differs from the value $\mu_P={\rm g}(n+\tilde{n}^0)$ obtained in the 
Popov approximation \cite{N2}. At $T=0$ this approximation gives 
$\mu_P(T=0)=4\pi\hbar^2an/m(1+8(a^3n)^{1/2}/3\sqrt{\pi})$, where the correction 
to ${\rm g}n$ comes from the quantum depletion of the condensate.
The $a^{5/2}$ order of the correction is correct {\it but} the numerical 
coefficient is wrong. To obtain the $T=0$ chemical potential to order $a^{5/2}$
one must use eq. (\ref{chempot}) and expand the coupling constant ${\rm g}$ 
to second order in the scattering length
\begin{equation}
{\rm g}=\frac{4\pi\hbar^2a}{m}\left(1+\frac{4\pi\hbar^2a}{m}\frac{1}{V}
\sum_{\bf p}\frac{m}{p^2} \right) \;\;.
\label{gren}
\end{equation}
The need of the above renormalization of ${\rm g}$ for the calculation of the 
ground-state energy of a Bose gas, has been first pointed out by Lee, Huang 
and Yang \cite{LHY} and is discussed in many textbooks
(see e.g. Ref. \cite{PAT}). 
By substituting the renormalized value of ${\rm g}$ into the first
term of eq. (\ref{chempot}) one gets
\begin{eqnarray}
\mu(T=0) &=& \frac{4\pi\hbar^2an}{m}\left(1+\frac{8}{3\sqrt{\pi}}(a^3n)^{1/2}
+\frac{4\pi\hbar^2a}{m}\frac{1}{V}\sum_{\bf p}\left( \frac{m}{p^2}-\frac{1}
{2\epsilon_p} \right) \right) \nonumber \\
&=& \frac{4\pi\hbar^2an}{m}\left( 1+ \frac{32}{3\sqrt{\pi}}(a^3n)^{1/2} \right)
\;\;,
\label{chempot1}
\end{eqnarray}     
which coincides with the well-known result for the chemical potential of 
a dilute Bose gas at 
$T=0$ \cite{LP1}.
In the above equation we have used (\ref{plwave})-(\ref{uv}) for the product 
$u_{\bf p}({\bf r})v_{\bf p}^{\ast}({\bf r})$ with the Bogoliubov spectrum 
(\ref{excen}). Notice that in the integral over ${\bf p}$ the ultraviolet 
divergencies arising from the renormalization of ${\rm g}$ and from 
$\tilde{m}^0$ cancel out. By differentiating (\ref{chempot1}) with respect 
to $n$ one obtains for the velocity of sound
\begin{equation}
c(T=0) = \sqrt{\frac{4\pi\hbar^2an}{m^2}}\left( 1+\frac{8}{\sqrt{\pi}}
(a^3n)^{1/2} \right) \;\;,
\label{svel}
\end{equation}
a result which has been first derived by Beliaev \cite{BEL}. In a 
self-consistent dynamic theory result (\ref{svel}) should also be obtained
from the frequency shift of the long-wavelength excitations. However, it 
can not be found within the Popov approximation because the crucial ingredients,
renormalization of ${\rm g}$ and equilibrium value of the anomalous density,
are not accounted for in this approximation.  

At finite $T$ one gets from eq. (\ref{chempot}) 
\begin{eqnarray}
\mu(T) &=&  
\frac{4\pi\hbar^2an}{m}\left( 1+ \frac{32}{3\sqrt{\pi}}\frac{n_0}{n}
(a^3n_0)^{1/2} \right) \nonumber \\
&+& \frac{4\pi\hbar^2an}{m} \frac{\sqrt{32}}{\sqrt{\pi}}
\frac{n_0}{n}(a^3n_0)^{1/2}\tau\int_0^\infty dx 
\frac{(\sqrt{1+\tau^2 x^2}-1)^{3/2}}{\sqrt{1+\tau^2 x^2}}\frac{1}
{e^x-1} \;\;,
\label{chempot2}
\end{eqnarray}
where $\tau=k_BT/mc^2$ is the reduced temperature. Result (\ref{chempot2}) 
coincides with the finding of Ref. \cite{SHI} obtained using the 
second-order Beliaev approximation. In the 
low-temperature 
regime, $\tau\ll 1$, one has  
\begin{equation}
\mu(T)=\mu(T=0)+\frac{\pi^2}{60}\frac{(k_BT)^4}{n\hbar^3 c^3} 
\label{chempot3}
\end{equation}
for the chemical potential, and
\begin{equation}
\kappa_T^{-1}=\rho c^2(T=0) - \frac{\pi^2}{24}\frac{(k_BT)^4}{\hbar^3c^3}
\label{invcomp1}
\end{equation}
for the inverse compressibility. The above results coincide with the ones 
obtained from the thermodynamic relation $\mu=(\partial F_{ph}/\partial N)
_{V,T}$, where $F_{ph}$ is the free energy of a phonon gas.  
In the same temperature regime, 
$k_BT\ll mc^2$, 
the Popov approximation gives $\mu_P(T)=\mu_P(T=0)+m^2c(k_BT)^2/(12n\hbar^3)$
and the inverse compressibility exhibits an unphysical $T^2$ dependence which
is not consistent with the $T^4$ dependence obtained by differentiating the 
phonon free energy.

In the low-temperature limit the frequency shift of collisionless phonons
is proportional to $T^4\log(T)$ \cite{AK,LP,PG}, whereas the Popov 
approximation 
again yields an unphysical $T^2$ dependence arising from 
the low-$T$ expansion of $n_0(T)$. 
The analogy with the result for the chemical 
potential suggests that also for this calculation  
the inclusion of $\tilde{m}^0$
is crucial to obtain the correct result.  

From the above results we conclude that a self-consistent dynamic theory, 
aiming to describe both the damping and the frequency shifts of the 
oscillations of the condensate, should go beyond the Popov ansatz 
(\ref{Popov}). The ingredients that this theory should contain 
are the following:
i) the equilibrium anomalous density $\tilde{m}^0$ has to be taken into 
account, ii) the renormalization (\ref{gren}) of the interaction coupling 
constant is needed in order to reproduce the energetics at $T=0$, iii) 
the elementary excitation energies $\epsilon_p$ have to be gapless and 
must coincide with the Bogoliubov spectrum at low-temperatures.

\section*{Acknowledgments}

The author wishes to thank L.P. Pitaevskii and S. Stringari for enlightning 
discussions and useful criticisms about the manuscript. Useful remarks made
by A. Griffin and F. Dalfovo are also gratefully acknowledged.

\begin{figure}
\caption{Dimensionless function $F$ as a function of $\tau$ (solid line). The 
asymptotic behaviours for $\tau\ll 1$ (dashed line), and for $\tau\gg 1$ 
(long-dashed line) are also reported.}
\end{figure}


\begin{references}
\bibitem{N1} Beliaev damping can also be calculated at finite temperature   
and it is given by $\gamma_B(T)=\gamma_B(T=0) n/n_0
(1+60\int_0^1 dx x^2(x-1)^2/(\exp(xcq/k_BT)-1))$. 
In the quantum regime ($k_BT\ll 
cq$) one has $\gamma_B(T)=\gamma_B(T=0)
(1+120(k_BT/cq)^3)$, while in the thermal regime ($k_BT\gg cq$)  
one finds
$\gamma_B(T)=5\gamma_B(T=0) k_BTn/(n_0cq)$ 
and the Beliaev damping is overwhelmed by Landau damping.
\bibitem{N2} For a discussion on the chemical potential in various 
approximations for Bose-condensed  systems and its connection to the 
Hugenholtz-Pines theorem see Ref. \cite{GRI}.
\end{references}
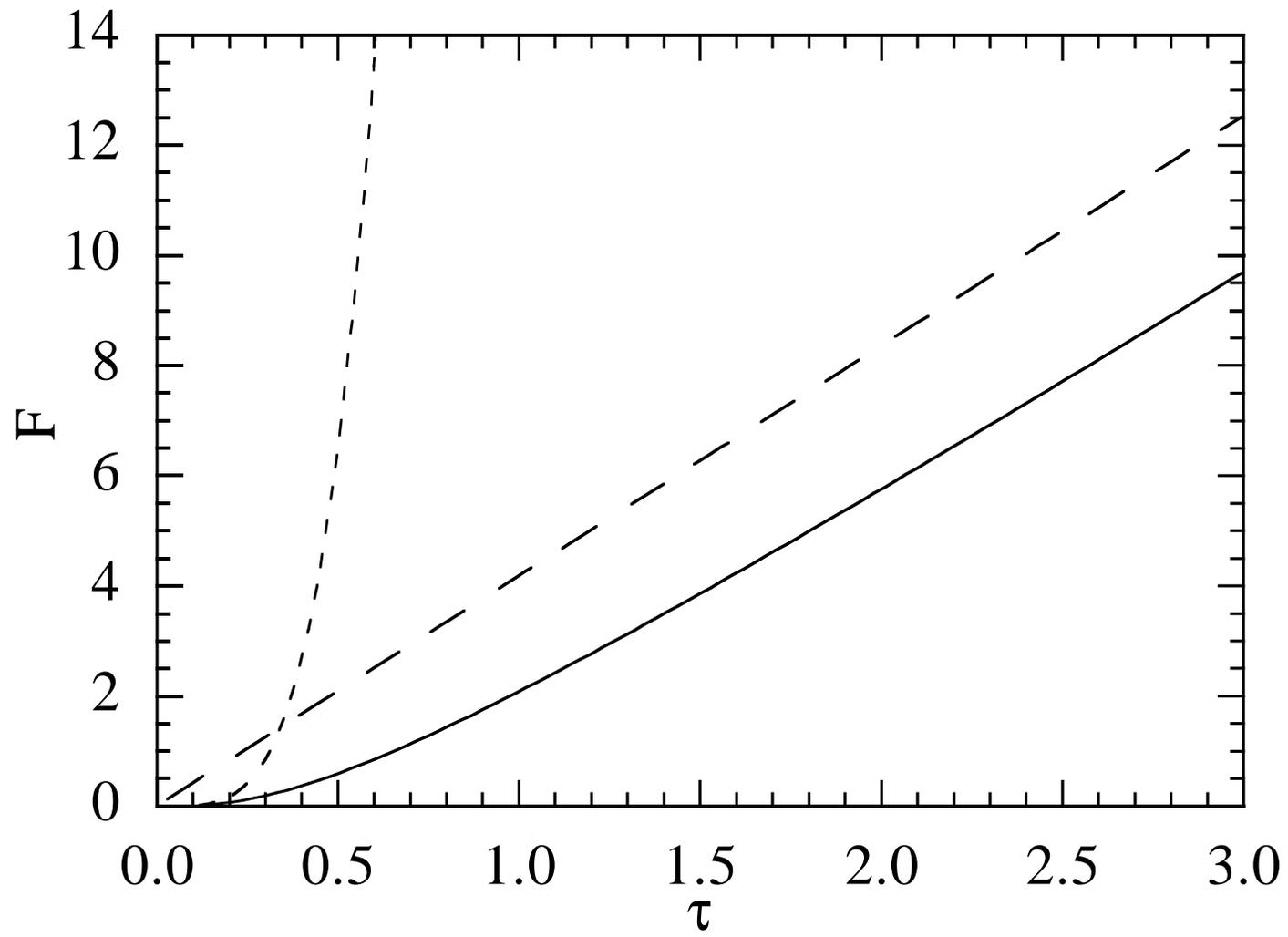
\end{document}